\documentclass[10pt,conference,final,twoside]{IEEEtran}
\bstctlcite{IEEEexample:BSTcontrol}
\usepackage{xspace,empheq,fancybox,amsmath,amssymb,graphicx,epstopdf,epsfig,syntonly,amsthm} 
\usepackage{bm}
\usepackage{url,cite,footnote,xspace,syntonly}
\usepackage{verbatim,multirow}
\usepackage[T1]{fontenc}
\usepackage{mathtools}
\usepackage{mwe}
\usepackage{algorithm}
\usepackage{algpseudocode}
\usepackage{graphicx}
\usepackage{textcomp}
\usepackage{xcolor}
\usepackage{multicol}
\usepackage{siunitx,booktabs}
\usepackage{csvsimple}
\allowdisplaybreaks

\newcommand{\review}[1]{\textcolor{black}{#1}}

\begin{document}
\title{Mitigating the Impact of Uncertain Wildfire Risk on Power Grids through Topology Control}
\author{Yuqi Zhou$^\dag$, Kaarthik Sundar$^\ddag$, Hao Zhu$^\dag$, and Deepjyoti Deka$^\ddag$ 
\thanks{
$^\dag$Department of Electrical Engineering, University of Texas, Austin, TX
}\;
\thanks{E-mail ids: \texttt{\{zhouyuqi,haozhu\}@utexas.edu}}\;
\thanks{
$^\ddag$Los Alamos National Laboratory, Los Alamos, NM
}\;
\thanks{E-mail ids: \texttt{\{kaarthik,deepjyoti\}@lanl.gov}}\;
\thanks{The authors acknowledge the funding provided by LANL’s Directed Research and Development (LDRD) project: ``Resilient operation of interdependent engineered networks and natural systems''. The research work conducted at Los Alamos National Laboratory is done under the auspices of the National Nuclear Security Administration of the U.S. Department of Energy under Contract No. 89233218CNA000001. The research work at UT-Austin has been supported by NSF under Awards 1802319, 2130706 and 2150571, as well as by PSERC. }
}

\IEEEoverridecommandlockouts

\maketitle

\begin{abstract}
Wildfires pose a significant threat to the safe and reliable operations of the electric grid. To mitigate wildfire risk, system operators resort to public safety power shutoffs, or PSPS, that shed load for a subset of customers. As wildfire risk forecasts are stochastic, such decision-making may often be sub-optimal. This paper proposes a two-stage topology control problem that jointly minimizes generation and load-shedding costs in the face of uncertain fire risk. Compared to existing work, we include pre- and post-event topology control actions and consider scenarios where the wildfire risk is known with low and high confidence. The effectiveness of the proposed approach is demonstrated using a benchmark test system, artificially geo-located in Southern California, and using stochastic wildfire risk data that exists in the literature. Our work provides a crucial study of the comparative benefits of pre-event versus post-event control and the effects of wildfire risk accuracy on each control strategy.
\end{abstract}

\begin{IEEEkeywords}
Topology control, Wildfire, line failures, stochastic optimization, Progressive Hedging
\end{IEEEkeywords}

%\newpage

%%%%%%%%%%%%%%%%%%%%%%%%%%%%%%%%%%%%%%%%%%%%%%%%%%%%%%%%%%%%%%%%%%%%%%
% %%
% %% Section: Intro %%
% %%%%%%%%%%%%%%%%%%%%%%%%%%%%%%%%%%%%%%%%%%%%%%%%%%%%%%%%%%%%%%%%%%%%
%%%

\section{Introduction}\label{sec:intro}
\review{Climate change has made effective wildfire mitigation a pressing challenge for reliable power system operations.} In the United States in 2021, there were over 58,000 wildfires that burned more than 7.13 million acres \cite{NIFC}. From 2019 to 2021, the number of wildfire cases increased by 16.9\%, and the number of acres burned increased by 51.1\%. Wildfires can be caused by lightning, arson, or power line faults termed high-impedance faults \cite{vazquez2022wildfire,li2021physics}. Additionally, a spreading wildfire's smoke can contaminate the insulating medium of transmission lines, leading to line outages and additional power shutoffs \cite{TR1}. A key distinguishing feature of wildfire-grid interaction is that, unlike other disasters, the grid can cause wildfires (sparks from line faults) and its effect (loss of load due to fire damage). As such, both prevention and mitigation of wildfire-related electricity disruption are crucial for grid resilience.

To prevent fires from originating and spreading from transmission lines, system operators de-energize lines through a practice known as Public Safety Power Shutoff (PSPS). As power shutoff can have an economic impact, several recent research has focused on designing control plans that minimize the risk of fire ignitions with a trade-off on load not served. The cost associated with load served can be modeled as maximizing load delivery \cite{coffrin2018relaxations,kody2022sharing,taylor2023managing}, minimizing system risks \cite{rhodes2020balancing,astudillo2022managing,kody2022optimizing,taylor2022framework}, and enhancing grid resilience \cite{trakas2017optimal,nazemi2022powering}. While these approaches are well-suited for de-energizing lines under a deterministic/known wildfire risk, they do not extend to the setting of uncertain wildfire risk that evolves over several days \cite{Cal}. Designing control policies for stochastic fire risk and associated PSPS policies will help quantify the effect that the accuracy of wildfire risk forecast has on resulting load shutoffs. 

\begin{figure}[ht!]
\centering
\includegraphics[width= .5\textwidth]{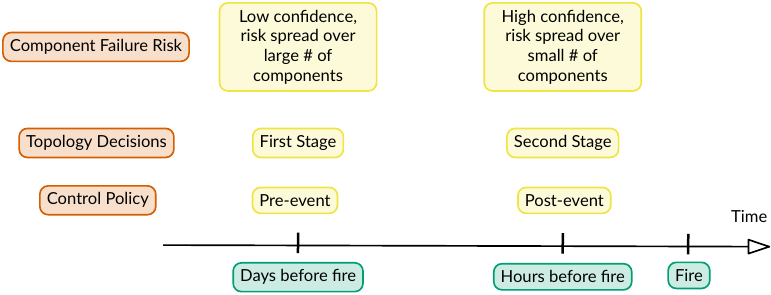}
\caption{Wildfire forecast timeline with control variables pre-event and post-event control policies.}
\label{fig:approach} 
\end{figure}
\begin{figure}[t!]
\centering
\includegraphics[trim=0cm 0cm 0cm 0cm,clip=true,totalheight=0.15\textheight]{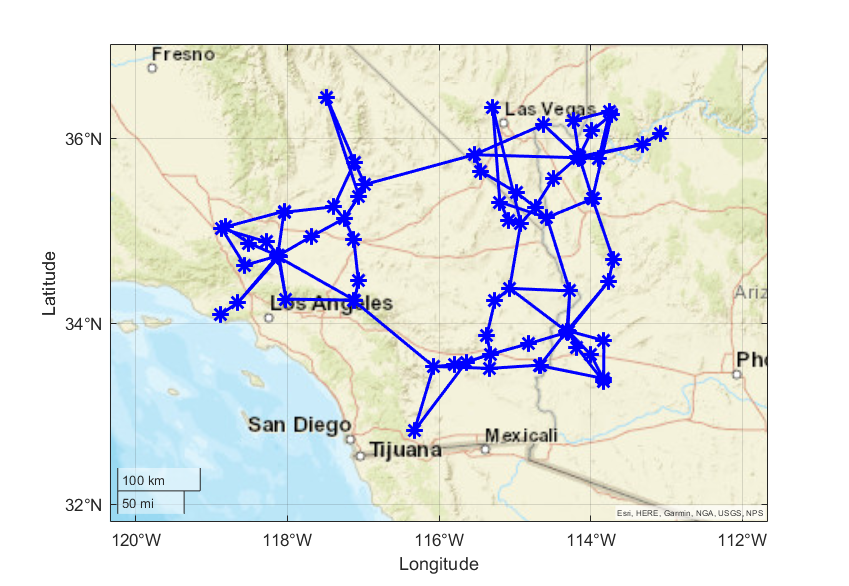}
\caption{RTS-GMLC test system with geographic information}
\label{fig:ca}
\end{figure}

\textbf{Contributions:} This paper aims to present pre and post-event topology control strategies to minimize average operating cost and load shed under uncertain wildfire risk. Each realization of the uncertainty wildfire risk model corresponds to a specific set of de-energized lines as mandated by the fire/grid operator's PSPS policy. The topology control design is modeled as a two-stage stochastic mixed-integer linear program for uncertain PSPS scenarios. It considers a linearized DC-OPF (optimal power flow) formulation with generation ramping, load shedding, and topology control as decision variables. \emph{`Pre-event'} control seeks to find an optimal network topology in advance to reduce the expected cost and load shed for all PSPS scenarios. The \emph{`post-event} control allows for different network configurations in the second stage of the stochastic program for each wildfire risk and corresponding PSPS scenario. This is highlighted in Fig.~\ref{fig:approach}. We solve both versions of the stochastic optimization problem using the \textbf{Progressive Hedging (PH)} algorithm \cite{watson2011progressive}. We demonstrate its scalability over alternate approaches through simulations on an RTS-GMLC test system, artificially geo-located in Southern California (see Fig.~\ref{fig:ca}), over two settings: one where the PSPS scenarios generated from the wildfire risk are more dispersed (low confidence forecast), and the other where the scenarios are concentrated in a smaller geographical region (more accurate forecast). Our simulation results show that, while less practical, post-event policies have lower system-wide costs than pre-event policies. However, optimal pre-event policies can perform similarly to optimal post-event actions in high-confidence settings where the wildfire risk is more concentrated.

%It is worth mentioning that our framework is generalizable to any candidate PSPS scenario(s) provided by the operator, that can include shut-offs to prevent fire ignition \cite{rhodes2020balancing}, or to prevent spread in neighboring lines, or multiple fires. 
 
The rest of the paper is organized as follows. Section \ref{sec:ps} presents the generation policy of scenarios for wildfires and the pre-event topology control formulation. The post-event control problem is formulated in Section \ref{sec:corrective}. Section \ref{sec:ph} presents an overview of the PH algorithm applied to both topology control formulations. Numerical experiments using the RTS-GMLC systems are presented in Section \ref{sec:results} to corroborate the effectiveness and efficiency of the proposed algorithms. Conclusions and possible avenues for future work of our work are presented in Section \ref{sec:con}. \\

\section{System Modeling}\label{sec:ps}
We consider a power grid $\mathcal{G}$ with a set of buses $\mathcal{N}$ and a set of lines $\mathcal{E}$. We first discuss the uncertain model for wildfire risk and corresponding PSPS scenarios. 
\subsection{Modeling scenarios using uncertain wildfire risk} \label{subsec:line-outages}
We consider a scenario-based uncertainty set $\mathcal{S}$ for wildfire risk, where each risk realization corresponds to a PSPS scenario $s$ of line-shutoffs $\xi_{ij}^s$. Here $\xi_{ij}^s$ takes a value $1$ or $0$ if line $(i, j) \in \mathcal E$ is operational or shut-off, respectively, in scenario $s$. We assume the maximum number of line shutoffs per scenario is bounded above by $\bm m$. We determine $\xi_{ij}^s$ for each line $(i, j)$ in scenario $s$ by sampling from a Bernoulli random variable (on/off) with a mean equal to the current estimate of the normalized risk of the line $(i, j)$. \footnote{We use data in \cite{rhodes2020balancing} for normalized risk estimates.} To model differing confidence in wildfire risk, we consider a risk-threshold $\bm R$ (detailed in Section \ref{subsec:test-system}) that restricts shutoffs in scenario $s$ to only lines, whose normalized risk value is greater than $\bm R$, where a higher $\bm R$ implies greater confidence in current risk. Thus, our uncertainty model has three parameters, number of scenarios $|S|$, number of shut-offs per scenario  $\bm m$, and risk confidence/threshold $\bm R$, that enable us to model fairly general PSPS policies. Next, we present the pre-event topology control problem to reduce cost and load-shed associated with the uncertain PSPS scenarios.

\begin{comment}
\textbf{Deterministic Topology Control}    
\end{comment}
\subsection{Pre-event Topology Control Formulation}
\label{sec:preventive}
As described in the Introduction, pre-event control determines an optimal topology over all PSPS scenario set $\mathcal S$, generated using Section \ref{subsec:line-outages}. We formulate this as a two-stage stochastic program. The first stage decisions, taken before the uncertainty (PSPS scenario) is realized, include the generation levels for each generator in the system and the topology decisions on the transmission lines. The second stage/ recourse decisions, made upon realization of PSPS, include ramping at each generator, load shed at each bus, and associated bus phase angles and line flows. The second stage decisions are a function of the specific scenario $s \in \mathcal S$. 

Let $p_i^g$ denote the real power generated at bus $i \in \mathcal N$. Associated with each line in $\mathcal E$, we introduce binary variables, $z_{ij}$ to model topology switching status. $z_{ij}$ takes a value $1$ when the line $(i, j) \in \mathcal E$ is closed and $0$, otherwise. The first stage decision variables are given by the set $\{ (p_i^g), (z_{ij})\}$.
Corresponding to each scenario $s \in \mathcal S$, we let $\theta_i^s$, $r_i^s$, and $\ell_i^s$ denote the voltage phase angle, the ramping, and the load shed at the bus $i \in \mathcal N$. Without loss of generality, we assume all generators can provide either up or down ramping. For each PSPS scenario $s \in \mathcal S$, we let $p_{ij}^s$ and $\theta_{ij}^s \triangleq \theta_i^s - \theta_j^s$ denote the active power flow and the phase angle difference for each line $(i, j) \in \mathcal E$. The second stage decision variables are summarized by the set $\{ (\theta_i^s), (r_i^s), (\ell_{i}^s), (p_{ij}^s), (\theta_{ij}^s)\}$. We assume that all the scenarios in the set $\mathcal S$ are equally likely, i.e., $\operatorname{Pr}(s) = \pi_s = |\mathcal S|^{-1}$. The pre-event topology control strategy is formulated as follows: 
\begin{subequations} \label{eq:preventive}
\begin{flalign}
& \min \quad \sum_{i \in \mathcal N} \bm c_i \cdot p_i^g + \mathbb E\left[ \sum_{i \in \mathcal N} \left( \bm c^r_i \cdot r_i^s + \review{\bm c^{\text{voll}}_i \cdot \ell_i^s} \right) \right] & \\ 
& \text{first stage variables \& constraints: } & \notag \\ 
& p_i^g \in [p_i^{gl}, p_i^{gu}] \quad \forall i \in \mathcal N & \label{eq:prev_1_gen_limits} \\
& z_{ij} \in \{0, 1\} \quad \forall (i, j) \in \mathcal E & \label{eq:prev_z} \\ 
% & z_{ij}^{\text{off}} \in \{0, 1\} \quad \forall (i, j) \in \mathcal E^{\text{off}}& \label{eq:prev_binary} \\
& \sum_{(i, j) \in \mathcal E} z_{ij} \leqslant\bm \beta & \label{eq:prev_budget} \\ 
& \text{second stage variables \& constraints $\forall s \in \mathcal S$: } & \notag \\
& \theta^s_{\bm r} = 0 \quad & \label{eq:prev_ref_bus} \\
& p_i^g + r_i^s \in [p_i^{gl}, p_i^{gu}] \quad \forall i \in \mathcal N & \label{eq:prev_2_gen_limits} \\
& \ell_i^s \in [0, \bm p_i^d] \quad \forall i \in \mathcal N & \label{eq:prev_ls} \\
& p_i^g + r_i^s - (\bm p_i^d - \ell_i^s) = \sum_{(i, j) \in \mathcal E_s} p_{ij}^s - \sum_{(j, i) \in \mathcal E_s} p_{ji}^s \quad \forall i \in \mathcal N & \label{eq:prev_kcl} \\ 
& p_{ij}^s \in [-\bm t_{ij}\xi_{ij}^s (1 - z_{ij}), \bm t_{ij} \xi_{ij}^s (1 - z_{ij})] ~ \forall (i, j) \in \mathcal E &\label{eq:prev_thermal_limits_on} \\
% & p_{ij}^s \in [-\bm t_{ij} \xi_{ij}^s z_{ij}^{\text{off}}, \bm t_{ij} \xi_{ij}^sz_{ij}^{\text{off}}] ~ \forall (i, j) \in \mathcal E^{\text{off}} &\label{eq:prev_thermal_limits_off} \\
& p_{ij}^s+ \bm b_{ij} \left\{ \theta_{ij}^s + \bm \theta \left(1-\xi_{ij}^s (1 - z_{ij}) \right) \right\} \geqslant 0 ~ \forall (i, j) \in \mathcal E & \label{eq:prev_pf_lb_on} \\ 
& p_{ij}^s+ \bm b_{ij} \left\{ \theta_{ij}^s - \bm \theta \left(1-\xi_{ij}^s (1 - z_{ij}) \right) \right\} \leqslant 0 ~ \forall (i, j) \in \mathcal E & \label{eq:prev_pf_ub_on} %\\ 
% & p_{ij}^s+ \bm b_{ij} \left\{ \theta_{ij}^s + \bm \theta \left(1-\xi_{ij}^s z_{ij}^{\text{off}} \right) \right\} \geqslant 0 \quad \forall (i, j) \in \mathcal E^{\text{off}} & \label{eq:prev_pf_lb_off} \\ 
% & p_{ij}^s+ \bm b_{ij} \left\{ \theta_{ij}^s - \bm \theta \left(1-\xi_{ij}^s z_{ij}^{\text{off}} \right) \right\} \leqslant 0 \quad \forall (i, j) \in \mathcal E^{\text{off}} & \label{eq:prev_pf_ub_off}
% & \theta_{ij} \geqslant -\bm \theta_{ij}^{\Delta} \cdot z_{ij} - \bm \theta \cdot (1 - z_{ij}) \quad \forall (i, j) \in \mathcal E & \label{eq:ots_theta_lb} \\ 
% & \theta_{ij} \leqslant \bm \theta_{ij}^{\Delta} \cdot z_{ij} + \bm \theta \cdot (1 - z_{ij}) \quad \forall (i, j) \in \mathcal E & \label{eq:ots_theta_ub} \\ 
% & \sum_{(i, j) \in \mathcal E} (1 - z_{ij}) \leqslant\bm \beta & \label{eq:ots_budget} \\ 
% & z_{ij} \in \{0, 1\} \quad \forall (i, j) \in \mathcal E & \label{eq:ots_binary}
\end{flalign}
\end{subequations}
The optimization problem in \eqref{eq:preventive} minimizes the sum of the nominal generation costs, the expected ramping costs, and the expected value of lost load over all PSPS scenarios. Under a budget of switching at most $\bm \beta$ lines, constraints in \eqref{eq:prev_1_gen_limits} -- \eqref{eq:prev_budget} represent the limits on the first stage variables. The constraints in \eqref{eq:prev_ref_bus} -- \eqref{eq:prev_kcl} enforce the operating limits for the second-stage variables and the post-PSPS DC power flow physics. Specifically, line switching decisions $z_{ij}$ are coupled with the line PSPS scenarios $\xi_{ij}^s$ to compute the topology status of line $(i, j)$ for scenario $s \in \mathcal S$. The thermal limits in \eqref{eq:prev_thermal_limits_on} and the DC power flow constraints in \eqref{eq:prev_pf_lb_on}, \eqref{eq:prev_pf_ub_on} for a line $(i, j) \in \mathcal E$ are enforced only under the following condition: $\xi_{ij}^s = 1$, $(1-z_{ij}) = 1$. T trivial bounds using big $\bm M$ constraints are enforced in all other cases. The next section describes the extension of the pre-event to post-event topology control problem.

\section{Post-event Topology Control Formulation}
\label{sec:corrective}
In contrast to the pre-event formulation, where topology control is the first-stage decision, the post-event control decides on the topology control actions in the recourse (or second-stage) problem. Performing topology control once the PSPS scenario is realized/known provides more flexibility. It can aid in achieving reduced ramping and load shed, though it might be less practical to implement. The formulation follows:
\begin{subequations} \label{eq:corrective}
\begin{flalign}
& \min \quad \sum_{i \in \mathcal N} \bm c_i \cdot p_i^g + \mathbb E\left[ \sum_{i \in \mathcal N} \left( \bm c^r_i \cdot r_i^s + \review{\bm c^{\text{voll}}_i \cdot \ell_i^s} \right) \right] & \\ 
& \text{first stage variables \& constraints: \eqref{eq:prev_1_gen_limits}} & \notag \\ 
& \text{second stage variables \& constraints $\forall s \in \mathcal S$:} & \notag \\
& \text{\eqref{eq:prev_ref_bus}, \eqref{eq:prev_2_gen_limits}, \eqref{eq:prev_ls}, \eqref{eq:prev_kcl}, and } & \notag \\
& z_{ij}^{s} \in \{0, 1\} \quad \forall (i, j) \in \mathcal E & \label{eq:corr_z_on} \\ 
% & z_{ij}^{\text{off},s} \in \{0, 1\} \quad \forall (i, j) \in \mathcal E^{\text{off}}& \label{eq:corr_binary} \\
& \sum_{(i, j) \in \mathcal E} z_{ij}^{s} \leqslant\bm \beta & \label{eq:corr_budget} \\ 
& p_{ij}^s \in [-\bm t_{ij}\xi_{ij}^s (1 - z_{ij}^{s}), \bm t_{ij} \xi_{ij}^s (1 - z_{ij}^{s})] ~ \forall (i, j) \in \mathcal E &\label{eq:corr_thermal_limits_on} \\
% & p_{ij}^s \in [-\bm t_{ij} \xi_{ij}^s z_{ij}^{\text{off},s}, \bm t_{ij} \xi_{ij}^sz_{ij}^{\text{off},s}] ~ \forall (i, j) \in \mathcal E^{\text{off}} &\label{eq:corr_thermal_limits_off} \\
& p_{ij}^s+ \bm b_{ij} \left\{ \theta_{ij}^s + \bm \theta \left(1-\xi_{ij}^s (1 - z_{ij}^{s}) \right) \right\} \geqslant 0 ~ \forall (i, j) \in \mathcal E & \label{eq:corr_pf_lb_on} \\ 
& p_{ij}^s+ \bm b_{ij} \left\{ \theta_{ij}^s - \bm \theta \left(1-\xi_{ij}^s (1 - z_{ij}^{s}) \right) \right\} \leqslant 0 ~ \forall (i, j) \in \mathcal E & \label{eq:corr_pf_ub_on} %\\ 
\end{flalign}
\end{subequations}
The first-stage and second-stage decision variables for the above formulation are given by the sets $\{(p_i^g)\}$, and $\{(z_{ij}^{s}), (\theta_i^s), (r_i^s), (\ell_{i}^s), (p_{ij}^s), (\theta_{ij}^s)\}$, respectively. For each PSPS scenario $s \in \mathcal S$, the definitions of $z_{ij}^{s}$ for $(i, j) \in \mathcal E$ 
%and $z_{ij}^{\text{off},s}$ for $(i, j) \in \mathcal E^{\text{off}}$ 
are similar to the corresponding ones in Section \ref{sec:preventive}. The only difference between pre-event and post-event control is that the former has binary decision variables for topology control in the first stage. In contrast, the latter has them in the second stage. Mathematically, solving the post-event control problem to global optimality is a more challenging task, and thus, there is a stronger need for heuristics. In the next section, we present an overview of our approach using the PH algorithm \cite{Rockafellar1991,watson2011progressive} to solve these optimization formulations.

\section{Progressive Hedging (PH) Algorithm} \label{sec:ph}
For ease of exposition, we first present an abstract formulation of a scenario-based two-stage stochastic program \cite{birge2011introduction} and use it to explain the different steps of the PH algorithm.  
\begin{comment}
    %
\begin{flalign}
\min_{\vec x \in X} \; c^{\intercal} \vec x + \mathbb{E}_{\Omega}\left[ Q(\vec x, \omega) \right] \text{ s.t. } A \vec x \geqslant b,\; \vec x \geqslant 0 \label{eq:2-stage}
\end{flalign}
where $Q(\vec x, \omega)$ defines the second stage problem as 
\begin{flalign}
Q(\vec x, \omega) = \min_{\vec y \in Y} \; q^{\intercal}_{\omega} \vec y \;\text{ s.t. } \; T_{\omega} \vec x + W \vec y \geqslant h_{\omega}, \; \vec y \geqslant 0 \label{eq:2-stage-2}
\end{flalign}
\end{comment}
Consider a set $\mathcal S$ of $n$ scenarios ($s = 1, \dots, n$, each of probability $\pi_1, \dots, \pi_n$) for the uncertainty realization (PSPS scenarios in our case). In the following problem,
\begin{subequations}
\begin{gather}
\min_{\vec x \in X, \vec y_s \in Y} \; c^{\intercal} \vec x + \sum_{s = 1}^n \pi_s q_s^{\intercal} \vec y_s\\ 
A \vec x \geqslant b  \\
T_s \vec x + W \vec y_s \geqslant h_s  \\ 
\vec x, \vec y_s \geqslant 0 
\end{gather}
\label{eq:extensive-form}
\end{subequations}
$X$ and $Y$, respectively, model the binary or continuous restrictions on the first and second-stage decision variables, $\vec x$ and $\vec y$. 
%decomposition approaches. In this paper, we leverage the PH algorithm proposed by Rockafellar and Wets \cite{Rockafellar1991,watson2011progressive} to solve both pre- and post-event control formulations. 
The main idea of the PH algorithm is to introduce first-stage decision variables $\vec x_s$ for each scenario $s$ and force them to be equal to $\vec x$. Using Lagrange variables and regularization, \eqref{eq:extensive-form} can be relaxed as:

\begin{comment}
i.e., 
\begin{subequations}
\begin{gather}
\min_{\vec x_s \in X, \vec y_s \in Y} \quad \sum_{s = 1}^n \pi_s \left( c^{\intercal} \vec x_s + q_s^{\intercal} \vec y_s\right)\\
\vec x_s = \vec x  \label{eq:non-anticipativity} \\
A \vec x_s \geqslant b \\ 
T_s \vec x_s + W \vec y_s \geqslant h_s \\ 
\vec x_s, \vec y_s \geqslant 0 
\end{gather}
\label{eq:ph-start}
\end{subequations}
\end{comment}
%The constraints in \eqref{eq:non-anticipativity} that enforce the consistency are called \textit{non-anticipative} because they make $\vec x_s$ independent of the scenario. These constraints are then dualized to obtain a regularized relaxation for each scenario $s$, as follows
\begin{subequations}
\begin{gather}
\min_{\vec x_s \in X, \vec y_s \in Y} \; c^{\intercal} \vec x_s + \pi_s q_s^{\intercal} \vec y_s + \rho_s(\vec x_s - \vec x) + \frac {\bm \gamma} 2 \| \vec x_s - \vec x\|^2\\
A \vec x_s \geqslant b \\ 
T_s \vec x_s + W \vec y_s \geqslant h_s \\ 
\vec x_s, \vec y_s \geqslant 0 
\end{gather}
\label{eq:ph-subproblem}
\end{subequations}
The PH algorithm, in each iteration $k$, generates new solutions $\vec x_s^k$ for every scenario $s$ and an \textit{implementable solution} $\vec x_k$, by aggregating $\vec x_s^k$ as $\vec x_k = \sum_{s = 1}^n \pi_s\vec x_s^k$. The Lagrange multipliers $\rho_s$ in \eqref{eq:ph-subproblem} are updated scenario-wise at each iteration $k$ using a penalty parameter $\bm \gamma$ as follows:
\begin{flalign}
\rho_s^{k+1} = \rho_s^k + \bm \gamma (\vec x_s^k - \vec x_k) \label{eq:rho-update}
\end{flalign}
\eqref{eq:rho-update} ensures that as the dual values converge, $\vec x_s^k$ become equal. The PH algorithm is terminated when both the primal and the dual gaps in \eqref{eq:termination} are below the pre-specified tolerances. 
\begin{subequations}
\begin{flalign}
\text{Primal gap: } \quad & \|\vec x_k - \vec x_{k-1}\|^2, ~~ \text{Dual gap:} \quad\sum_{s = 1}^n \pi_s \| \vec x_s^k - \vec x_k \|^s
\end{flalign}
\label{eq:termination}
\end{subequations}
The PH algorithm, in comparison with alternatives like Bender's \cite{rahmaniani2017benders} or dual decomposition \cite{caroe1999dual}, is highly \textit{parallelizable} for both pre-event and post-event topology control. It possesses theoretical convergence guarantees\cite{biel2022efficient} and robust heuristics \cite{Fan2010,Listes2005,Lokketangen1996,biel2022efficient}. A pseudo-code of the algorithm is shown in Algorithm \ref{algo:ph}. 

\begin{algorithm}
\caption{Progressive Hedging: a pseudo-code}
\label{algo:ph}{
\begin{algorithmic}[1] % The number tells where the line numbering should start
\Statex \textbf{Initialization:} 
\State $k \gets 0$ \Comment{iteration count}
\State \label{step:main0}For every $s \in \mathcal S$, initialize $\vec x_{s}^k \gets\;$ solution of \eqref{eq:ph-subproblem} without the penalty terms $\rho_s(\vec x_s - \vec x) + \frac{\bm \gamma}2 \| \vec x_s - \vec x\|^2$
\State $\vec x_k = \sum_{s = 1}^n \pi_s \vec x_s^k$
\State $\rho_k \gets\;$ initial penalty parameter 
\Statex \textbf{Iteration update:} 
\State \label{step:goto} $k \gets k+1$ 
\Statex \textbf{Decomposition:}
\State \label{step:main}For every $s \in \mathcal S$, update $\vec x_{s}^k \gets \text{ Solution of \eqref{eq:ph-subproblem}}$ 
\Statex \textbf{Aggregation:}
\State$\vec x_k = \sum_{s = 1}^n \pi_s \vec x_s^k$
\Statex \textbf{Lagrangian multiplier update:}
\State For every $s \in \mathcal S$, $\rho_s^{k+1} = \rho_s^k + \bm \gamma (\vec x_s^k - \vec x_k)$
\Statex \textbf{Termination criterion check:}
\State $\epsilon_{p}^k \triangleq  \|\vec x_k - \vec x_{k-1}\|^2, \epsilon_{d}^k \triangleq \sum_{s = 1}^n \pi_s \| \vec x_s^k - \vec x_k \|^s$
\If{$\epsilon_{p}^k > \varepsilon_{p}$ or $\epsilon_{d}^k > \varepsilon_{d}$} 
\State Go to Step \ref{step:goto} 
\Else
\State Terminate with $\vec x^k$ as the first stage solution
\EndIf
\end{algorithmic}
}
\end{algorithm}

\section{Numerical Results}
\label{sec:results}
In this section, we present extensive computational experiments that corroborate the effectiveness of the proposed formulations and algorithms. We first start with briefly describing the test system and data used.

\subsection{Test System and Scenario Generation} \label{subsec:test-system} 
We use the RTS-GMLC system \cite{barrows2019ieee}, artificially geo-located in the southwestern part of the United States. This synthetic system consists of 73 buses and 120 transmission lines. We consider both load factors of $1$ (base case) and $1.05$ (higher loading). The linear cost coefficient for each generator in the test case is directly used for $\bm c_i$ in Problems \eqref{eq:preventive}, \eqref{eq:corrective}. The ramping and value of lost load's cost coefficients are set to $\bm c_i^r = 10\% \times \bm c_i$, and $\bm c_i^{\mathrm{voll}} = 10.0 \times \max_i(c_i)$, respectively. In our formulation, when a generator ramps up, we cost the excess generation using $\bm c_i+\bm c_i^r$, whereas we cost down ramping using only $\bm c_i^r$. This is because ramping generators up should be more expensive than ramping them down. In the PH Algorithm \ref{algo:ph}, the primal and dual tolerances for the termination criteria are set to $10^{-3}$ and $10^{-2}$, respectively. Finally, the number of lines that can be controlled ($\bm \beta$) is set to 5 for all the computational experiments.

\begin{figure}[t!]
\centering
\includegraphics[trim=0cm 0cm 0cm 0cm,clip=true,totalheight=0.20\textheight]{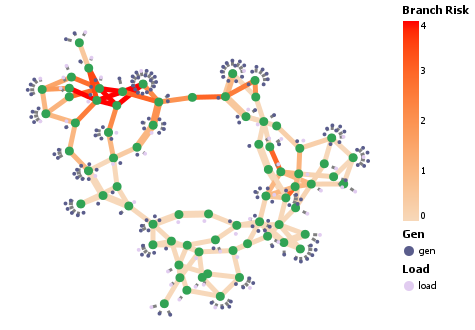}
\caption{\review{Heat map of line (branch) risk values for the RTS-GMLC system}}
\label{fig:risk}
\end{figure}

For the PSPS scenario generation procedure described in Section \ref{subsec:line-outages}, we utilize line failure risk values generated in \cite{rhodes2020balancing} and depicted in Fig.~\ref{fig:risk}. These risk values are transformed into probabilities of line failures using a combination of normalization and thresholding. $n$ PSPS scenarios are then generated with a maximum of $\bm m$ line-shotoffs each, by sampling. In all our computational experiments the value of $\bm m$ is set to $4$. The pseudo-code for generating scenarios is given in Algorithm \ref{algo:scenario-generation}. Note that a higher risk threshold (higher confidence in current risk) restricts the PSPS scenarios to fewer lines. 

\begin{algorithm}
\caption{PSPS scenario generation}
\label{algo:scenario-generation}{
\begin{algorithmic}[1] %
\Statex \textbf{Given:} 
\State \label{step:threshold}$\bm R \gets$ risk threshold
\State Wildfire risk for each line from \cite{rhodes2020balancing} - ${risk}_{ij}\; \forall (i, j) \in \mathcal E$
\State $n \gets$ no. of scenarios, $\bm m \gets$ max. no. of PSPS per scenario
\Statex \textbf{Approach:} 
\State $\mathcal L \gets \{(i, j) \in \mathcal E: \mathrm{risk}_{ij} \geqslant \bm R\}$, $\mathrm{risk}_{ij} \gets 0\; \forall (i, j) \in \mathcal E \setminus \mathcal L$
\State Normalize $\mathrm{risk}_{ij}$ to sum to one over all $(i, j) \in \mathcal E$
\State \review{Sample with replacement, using $\mathrm{risk}_{ij}$ as probabilities, to generate at most $\bm m$ shutoffs. Repeat to generate $n$ scenarios.
\State $\mathcal S \gets$ the $n$ shutoff scenarios 
\State Probability $\pi_s \gets \frac 1n \; \forall s \in \mathcal S$}
\end{algorithmic}
}
\end{algorithm}

\subsection{Computational Platform \& Implementation Details} \label{subsec:specs} 
All the presented formulations were implemented in the Julia Programming language \cite{bezanson2017julia} using JuMP \cite{DunningHuchetteLubin2017} as the mathematical programming layer and CPLEX as the MILP solver. All computational experiments were run on a regular laptop equipped with Intel\textsuperscript{\textregistered} CPU @ 2.60 GHz and 12 GB of RAM. \texttt{PowerModels.jl} \cite{8442948} was used to parse the RTS-GMLC's MATPOWER case file.

\textbf{Algorithm Performance:}
We first compare the computation time taken for the pre-event and post-event control when the sub-problems in Step \ref{step:main} of PH Algorithm \ref{algo:ph} are solved in serial v/s in parallel. Table \ref{tab:time-1} and \ref{tab:time-105} show the computation time with increasing scenario set $\mathcal S$, for load scaling of $1$ and $1.05$, respectively. For these runs, the risk threshold in Algorithm \ref{algo:scenario-generation} was set to $\bm R = 0$ (lowest confidence in the forecast), corresponding to all lines being candidates for PSPS.  The parallel version of the PH algorithm is always preferable to the serial version. Solving the post-event control formulation is three to seven times faster than the pre-event control formulation, as the former doesn't have binary variables in the first stage. The parallel version of the PH algorithm converged within $35$ iterations for pre and post-event control for every size of $\mathcal S$ considered. This outscored our simulation runs with Bender's \cite{rahmaniani2017benders} decomposition, which achieved an optimality gap of less than $10\%$ for only $2$ runs within the 3-hour computation time limit, for the pre-event setting.
\begin{table}[!htb]
\centering
\caption{Computation time in sec. and number of iterations (load scaling - $1.0$). Here, ``s-time'' is the time taken for the serial run, ``p-time'' is the time taken for the parallel run, and ``iter'' is the number of iterations for termination.}
\label{tab:time-1}
\begin{tabular}{ccccccc}
\toprule
\multirow{2}{*}{$|\mathcal S|$} & \multicolumn{3}{c}{pre-event} & \multicolumn{3}{c}{post-event} \\
\cmidrule{2-7}
& s-time & p-time & iter & s-time & p-time & iter \\ 
\midrule
\begin{filecontents*}{times-iter.csv}
scenarios,preventive-s-a,corrective-s-a,preventive-p-a,corrective-p-a,preventive-s-b,corrective-s-b,preventive-p-b,corrective-p-b,prev_iter_a,corr_iter_a,prev_iter_b,corr_iter_b
40,461.81,250.71,304.87,174.09,7910.58,916.99,8014.49,780.86,11,3,13,9
80,797.54,770.39,1274.08,308.85,8116.71,1407.84,8389.74,1541.47,11,3,12,10
120,2589,394.4,1772.02,438.45,9388.02,1937.38,8586.4,720.36,35,28,35,28
160,4080.08,982.4,2934.43,737.85,11272.08,2675.7,9025.12,1355.26,10,4,11,7
200,4931.17,689.13,3659.57,719.57,12487.13,2766.56,11076.21,1919.78,10,3,11,7
\end{filecontents*}
\csvreader[late after line=\\]{times-iter.csv}{
1=\one,2=\two,3=\three,4=\four,5=\five,6=\six,7=\seven,
8=\eight,9=\nine,10=\ten,11=\eleven,12=\twelve,13=\thirteen
}{\one & \two & \four & \ten & \three & \five & \eleven}
\bottomrule
\end{tabular}
\end{table}

\begin{figure}
\centering
\includegraphics[scale=0.8]{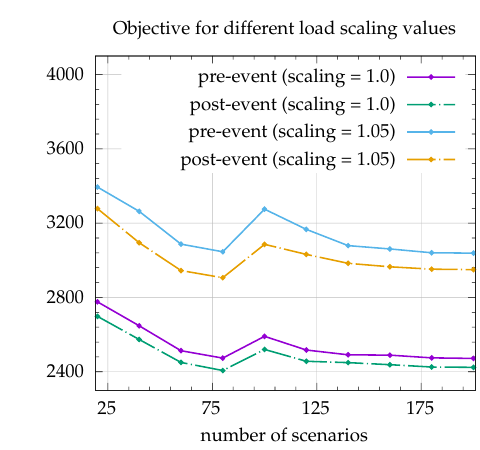}
\caption{Objective value of the pre-event and post-event control problems with varying scenarios for a load scaling of $1.0$ and $1.05$.}
\label{fig:obj}
\end{figure}

%\begin{comment}
\begin{table}[!htb]
\centering
\caption{Computation time in sec. and number of iterations (load scaling - $1.05$). Here, ``s-time'' is the time taken for the serial run, ``p-time'' is the time taken for the parallel run, and ``iter'' is the number of iterations for termination.}
\label{tab:time-105}
\begin{tabular}{ccccccc}
\toprule
\multirow{2}{*}{$|\mathcal S|$} & \multicolumn{3}{c}{pre-event} & \multicolumn{3}{c}{post-event} \\
\cmidrule{2-7}
& s-time & p-time & iter & s-time & p-time & iter \\ 
\midrule
\csvreader[late after line=\\]{times-iter.csv}{
1=\one,2=\two,3=\three,4=\four,5=\five,6=\six,7=\seven,
8=\eight,9=\nine,10=\ten,11=\eleven,12=\twelve,13=\thirteen
}{\one & \six & \eight & \twelve & \seven & \nine & \thirteen}
\bottomrule
\end{tabular}
\end{table}
\textbf{Comparison of solutions:} Fig.~\ref{fig:obj} compare the objective value obtained using the pre-event and the post-event control formulations for a varying number of scenarios. The objective values stabilize and converge to a limit with increasing scenarios. We also remark that, in the base case with no scenarios, when the problem in Sec.~\ref{sec:ps} was solved optimally, the total generation output was $85.5$ p.u. and no load-shedding was required. 

\begin{figure}[htbp]
\centering
\includegraphics[trim=0cm 0cm 0cm 0cm,clip=true,totalheight=0.20\textheight]{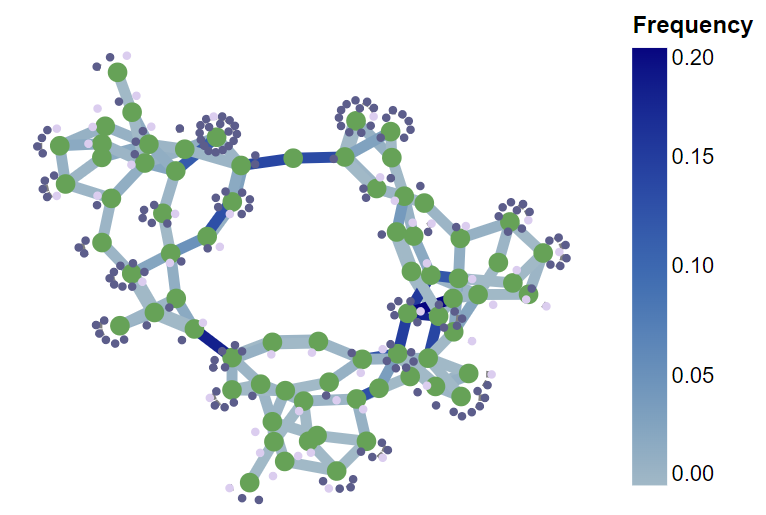}
\caption{\review{Frequency of switching for different lines under post-event control.}}
\label{fig:frequency}
\end{figure}

% \begin{figure}
% \centering
% \includegraphics{obj-105.pdf}
% \caption{Objective value of the pre-event and post-event control problems with varying scenarios for a load scaling of $1.05$.}
% \label{fig:obj-105}
% \end{figure}
Observe that the objective value of the pre-event control problem is always higher than that of the post-event control problem. Table \ref{tab:ls} shows the expected load shed over all the PSPS scenarios by the pre-event and the post-event control formulations. These results indicate that it is always prudent to choose the post-event control formulation to minimize the load shed; the pre-event switching policy is more practical as it can be done ahead of time and incurs no real-time changes. 
\begin{comment}
\begin{table}[!htb]
\centering
\caption{Relative gap between the post-event and pre-event formulations' objective values (relative w.r.t pre-event's objective value)}
\label{tab:rel-gap}
\begin{tabular}{ccc}
\toprule
$|\mathcal S|$ & \multicolumn{2}{c}{relative gap (\%)} \\
\cmidrule{2-3}
& load scaling - $1.0$ & load scaling - $1.05$ \\ 
\midrule
\csvreader[late after line=\\]{obj.csv}{
1=\one,2=\two,3=\three,4=\four,5=\five,6=\six,7=\seven
}{\one & \four & \seven}
\bottomrule
\end{tabular}
\end{table}
\end{comment}
Finally, Fig. \ref{fig:frequency} shows the ratio of the number of times a line was switched under the post-event control formulation to the total number of scenarios, which in this case was 200. When examined together with Fig. \ref{fig:risk}, it can be observed that the lines that get switched have a geographical correlation to the lines that have a wildfire risk. \\
\textbf{AC feasibility:} For each formulation (pre and post-event), the optimal topology under each PSPS scenario was checked for feasible AC power flow and was found to be feasible.

\begin{table}[!htbp]
\centering
\caption{Average load shed (in MW) over all the scenarios by the solutions of the pre-event and post-event control formulations}
\label{tab:ls}
\begin{tabular}{ccccc}
\toprule
\multirow{2}{*}{$|\mathcal S|$} & \multicolumn{2}{c}{load scaling - 1.0} & \multicolumn{2}{c}{load scaling - 1.05} \\
\cmidrule{2-5}
& pre-event & post-event & pre-event & post-event \\ 
\midrule
\begin{filecontents*}{ls-mean-reduced.csv}
numscenarios,pa,ca,pb,cb
40,33.48,29.72,52.01,41.63
80,22.48,18.93,37.19,28.64
120,20.43,17.32,37.78,28.44
160,23.44,20.74,37.96,32.21
200,22.39,19.98,36.49,31.15
\end{filecontents*}
\csvreader[late after line=\\]{ls-mean-reduced.csv}{
1=\one,2=\two,3=\three,4=\four,5=\five
}{\one & \two & \three & \four & \five}
\bottomrule
\end{tabular}
\end{table}

% \textcolor{red}{Deep: will put this figure with the figure of risk.. as subfigure}

\begin{figure}
\centering
\includegraphics[scale=0.8]{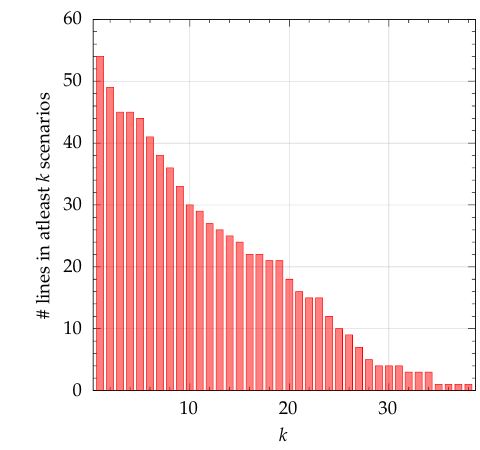}
\caption{Histogram of the PSPS scenarios generated using $\bm R = 0$. The shutoffs are spread over $54$ lines, indicating low confidence in the PSPS scenarios. The first bar in the histogram indicates that $54$ unique lines are contained in at least $1$ PSPS scenario.}
\label{fig:hist-no-threshold}
\end{figure}

\textbf{Impact of Risk Threshold:} All results presented thus far are on PSPS scenarios generated with $\bm R = 0$ (low confidence on risk), and keeping the value of $\bm m = 4$ in Algorithm \ref{algo:scenario-generation}. The histogram in Fig.~\ref{fig:hist-no-threshold} presents the number of lines in at least $k$ PSPS scenarios. It is clear that when $\bm R = 0$, PSPS scenarios are spread over most lines in the RTS-GMLC network, leading to a higher cost, particularly for pre-event control.
To understand the impact of confident risk values, we consider pre-and post-event formulations for a new set of $200$ PSPS scenarios limited to choosing shutoffs among $4$ highest risk lines. In this case, the load shed under both control formulations was observed to be relatively close to each other (see Table \ref{tab:ls-high}). This is intuitive as the number of possible scenarios has a high order of overlap. As such, when the wildfire risk is known with high confidence, pre-event control should be the right way forward as it is more easily implementable and does not lead to any significant loss in cost. For low-risk confidence, our method quantifies the optimality loss in pre-event (over post-event control). 
\begin{comment}
\begin{figure}
\centering
\includegraphics{scenarios-threshold-4.pdf}
\caption{Histogram when the risk is spread over $4$ lines. These four lines are present in at least $122$ out of the $200$ scenarios generated. The histogram indicates high confidence in the PSPS scenarios.}
\label{fig:hist-threshold}
\end{figure}
\end{comment}
\begin{table}
\centering
\caption{Average load shed (in MW) over all the scenarios for $200$ high confidence damage scenarios}
\label{tab:ls-high}
\begin{tabular}{cccc}
\toprule
\multicolumn{2}{c}{load scaling - 1.0} & \multicolumn{2}{c}{load scaling - 1.05} \\ 
\cmidrule(lr){1-4} 
pre-event & post-event & pre-event & post-event \\ 
\midrule 
\begin{filecontents*}{ls-mean-threshold.csv}
numscenarios,pa,ca,pb,cb
50,43.79,40.79,67.49,65.88
\end{filecontents*}
\csvreader[late after line=\\]{ls-mean-threshold.csv}{1=\s,2=\pa,3=\ca,4=\pb,5=\cb}{\pa & \ca & \pb & \cb}
\bottomrule
\end{tabular}
\end{table}
\section{Conclusions} \label{sec:con}
This paper considered the optimal topology control problem under line shutoffs due to uncertain wildfires. We proposed a scalable two-stage stochastic mixed-integer formulation that minimizes cost and load shed. Numerical studies on the practical RTS-GMLC system corroborate the performance of the proposed topology control algorithms and demonstrate that lower risk confidence can lead to a reduction in optimal operation of pre-event over post-event control policies.

This work leads to several directions of work. We are currently analyzing methods to ensure fair control policies under stochastic wildfires. Further, improved sampling strategies for wildfire scenarios (including ones based on active/importance sampling) will significantly enhance the current naive sampling strategy. 

\bibliography{bibliography}

% Generated by IEEEtran.bst, version: 1.14 (2015/08/26)
\begin{thebibliography}{10}
\providecommand{\url}[1]{#1}
\csname url@samestyle\endcsname
\providecommand{\newblock}{\relax}
\providecommand{\bibinfo}[2]{#2}
\providecommand{\BIBentrySTDinterwordspacing}{\spaceskip=0pt\relax}
\providecommand{\BIBentryALTinterwordstretchfactor}{4}
\providecommand{\BIBentryALTinterwordspacing}{\spaceskip=\fontdimen2\font plus
\BIBentryALTinterwordstretchfactor\fontdimen3\font minus \fontdimen4\font\relax}
\providecommand{\BIBforeignlanguage}[2]{{%
\expandafter\ifx\csname l@#1\endcsname\relax
\typeout{** WARNING: IEEEtran.bst: No hyphenation pattern has been}%
\typeout{** loaded for the language `#1'. Using the pattern for}%
\typeout{** the default language instead.}%
\else
\language=\csname l@#1\endcsname
\fi
#2}}
\providecommand{\BIBdecl}{\relax}
\BIBdecl

\bibitem{NIFC}
\BIBentryALTinterwordspacing
The {N}ational {I}nteragency {F}ire {C}enter. [Online]. Available: \url{https://www.nifc.gov/fire-information}
\BIBentrySTDinterwordspacing

\bibitem{vazquez2022wildfire}
D.~A.~Z. Vazquez, F.~Qiu, N.~Fan, and K.~Sharp, ``Wildfire mitigation plans in power systems: A literature review,'' \emph{IEEE Transactions on Power Systems}, 2022.

\bibitem{li2021physics}
W.~Li and D.~Deka, ``Physics-informed learning for high impedance faults detection,'' in \emph{2021 IEEE Madrid PowerTech}.\hskip 1em plus 0.5em minus 0.4em\relax IEEE, 2021, pp. 1--6.

\bibitem{TR1}
``Effect of wildfires on transmission line reliability,'' California Public Utilities Commission, Tech. Rep., 2008.

\bibitem{coffrin2018relaxations}
C.~Coffrin, R.~Bent, B.~Tasseff, K.~Sundar, and S.~Backhaus, ``Relaxations of ac maximal load delivery for severe contingency analysis,'' \emph{IEEE Transactions on Power Systems}, vol.~34, no.~2, pp. 1450--1458, 2018.

\bibitem{kody2022sharing}
A.~Kody, A.~West, and D.~K. Molzahn, ``Sharing the load: Considering fairness in de-energization scheduling to mitigate wildfire ignition risk using rolling optimization,'' \emph{arXiv preprint arXiv:2204.06543}, 2022.

\bibitem{taylor2023managing}
S.~Taylor, G.~Setyawan, B.~Cui, A.~Zamzam, and L.~A. Roald, ``Managing wildfire risk and promoting equity through optimal configuration of networked microgrids,'' in \emph{Proceedings of the 14th ACM International Conference on Future Energy Systems}, 2023, pp. 189--199.

\bibitem{rhodes2020balancing}
N.~Rhodes, L.~Ntaimo, and L.~Roald, ``Balancing wildfire risk and power outages through optimized power shut-offs,'' \emph{IEEE Transactions on Power Systems}, vol.~36, no.~4, pp. 3118--3128, 2020.

\bibitem{astudillo2022managing}
A.~Astudillo, B.~Cui, and A.~S. Zamzam, ``Managing power systems-induced wildfire risks using optimal scheduled shutoffs,'' in \emph{2022 IEEE Power \& Energy Society General Meeting (PESGM)}.\hskip 1em plus 0.5em minus 0.4em\relax IEEE, 2022, pp. 1--5.

\bibitem{kody2022optimizing}
A.~Kody, R.~Piansky, and D.~K. Molzahn, ``Optimizing transmission infrastructure investments to support line de-energization for mitigating wildfire ignition risk,'' \emph{arXiv preprint arXiv:2203.10176}, 2022.

\bibitem{taylor2022framework}
S.~Taylor and L.~A. Roald, ``A framework for risk assessment and optimal line upgrade selection to mitigate wildfire risk,'' \emph{Electric Power Systems Research}, vol. 213, p. 108592, 2022.

\bibitem{trakas2017optimal}
D.~N. Trakas and N.~D. Hatziargyriou, ``Optimal distribution system operation for enhancing resilience against wildfires,'' \emph{IEEE Transactions on Power Systems}, vol.~33, no.~2, pp. 2260--2271, 2017.

\bibitem{nazemi2022powering}
M.~Nazemi and P.~Dehghanian, ``Powering through wildfires: An integrated solution for enhanced safety and resilience in power grids,'' \emph{IEEE Transactions on Industry Applications}, vol.~58, no.~3, pp. 4192--4202, 2022.

\bibitem{Cal}
\BIBentryALTinterwordspacing
The {C}alifornia {D}epartment of {F}orestry and {F}ire {P}rotection ({CAL} {FIRE}). [Online]. Available: \url{https://www.fire.ca.gov/incidents}
\BIBentrySTDinterwordspacing

\bibitem{watson2011progressive}
J.-P. Watson and D.~L. Woodruff, ``Progressive hedging innovations for a class of stochastic mixed-integer resource allocation problems,'' \emph{Computational Management Science}, vol.~8, no.~4, p. 355, 2011.

\bibitem{Rockafellar1991}
R.~T. Rockafellar and R.~J.-B. Wets, ``Scenarios and policy aggregation in optimization under uncertainty,'' \emph{Mathematics of operations research}, vol.~16, no.~1, pp. 119--147, 1991.

\bibitem{birge2011introduction}
J.~R. Birge and F.~Louveaux, \emph{Introduction to stochastic programming}.\hskip 1em plus 0.5em minus 0.4em\relax Springer Science \& Business Media, 2011.

\bibitem{rahmaniani2017benders}
R.~Rahmaniani, T.~G. Crainic, M.~Gendreau, and W.~Rei, ``The benders decomposition algorithm: A literature review,'' \emph{European Journal of Operational Research}, vol. 259, no.~3, pp. 801--817, 2017.

\bibitem{caroe1999dual}
C.~C. Car{\o}e and R.~Schultz, ``Dual decomposition in stochastic integer programming,'' \emph{Operations Research Letters}, vol.~24, no. 1-2, pp. 37--45, 1999.

\bibitem{biel2022efficient}
M.~Biel and M.~Johansson, ``Efficient stochastic programming in julia,'' \emph{INFORMS Journal on Computing}, vol.~34, no.~4, pp. 1885--1902, 2022.

\bibitem{Fan2010}
Y.~Fan and C.~Liu, ``Solving stochastic transportation network protection problems using the progressive hedging-based method,'' \emph{Networks and Spatial Economics}, vol.~10, no.~2, pp. 193--208, 2010.

\bibitem{Listes2005}
O.~Listes and R.~Dekker, ``A scenario aggregation--based approach for determining a robust airline fleet composition for dynamic capacity allocation,'' \emph{Transportation Science}, vol.~39, no.~3, pp. 367--382, 2005.

\bibitem{Lokketangen1996}
A.~L{\o}kketangen and D.~L. Woodruff, ``Progressive hedging and tabu search applied to mixed integer (0, 1) multistage stochastic programming,'' \emph{Journal of Heuristics}, vol.~2, no.~2, pp. 111--128, 1996.

\bibitem{barrows2019ieee}
C.~Barrows, A.~Bloom, A.~Ehlen, J.~Ik{\"a}heimo, J.~Jorgenson, D.~Krishnamurthy, J.~Lau, B.~McBennett, M.~O’Connell, E.~Preston \emph{et~al.}, ``The ieee reliability test system: A proposed 2019 update,'' \emph{IEEE Transactions on Power Systems}, vol.~35, no.~1, pp. 119--127, 2019.

\bibitem{bezanson2017julia}
J.~Bezanson, A.~Edelman, S.~Karpinski, and V.~B. Shah, ``Julia: A fresh approach to numerical computing,'' \emph{SIAM review}, vol.~59, no.~1, pp. 65--98, 2017.

\bibitem{DunningHuchetteLubin2017}
I.~Dunning, J.~Huchette, and M.~Lubin, ``Jump: A modeling language for mathematical optimization,'' \emph{SIAM Review}, vol.~59, no.~2, pp. 295--320, 2017.

\bibitem{8442948}
C.~Coffrin, R.~Bent, K.~Sundar, Y.~Ng, and M.~Lubin, ``Powermodels.jl: An open-source framework for exploring power flow formulations,'' in \emph{2018 Power Systems Computation Conference (PSCC)}, June 2018, pp. 1--8.

\end{thebibliography}

\bibliographystyle{IEEEtran}
\end{document}